\documentclass[useAMS,usenatbib, usegraphicx]{mn2e}
\usepackage{amssymb, hyperref, rotating, multirow, threeparttable, mathptmx}
  \def\apjs{ApJS} 
\def\apjl{ApJ}   \def\aap{A\&A}
   
\let\ga=\gtrsim \let\la=\lesssim
\def\kms{km~s$^{-1}$}     
\def\Msun{M$_{\sun}$}     

\title[Delayed star-formation due to inflows] 
{Delayed star-formation in high-redshift stream-fed galaxies}

\author[Gabor et al.]{
J. M. Gabor,$^{1}$\thanks{Email:jared.gabor@cea.fr}
Fr\'ed\'eric Bournaud$^{1}$ 
\\ $^{1}$CEA-Saclay, IRFU, SAp, F-91191 Gif-sur-Yvette, France  
}

\begin{document}


\pagerange{\pageref{firstpage}--\pageref{lastpage}} \pubyear{2013}

\maketitle
\label{firstpage}

\begin{abstract}
We propose that star formation is delayed relative to the inflow rate
in rapidly-accreting galaxies at very high redshift ($z>2$) because of
the energy conveyed by the accreting gas. Accreting gas streams
provide fuel for star formation, but they stir the disk and increase
turbulence above the usual levels compatible with gravitational
instability, reducing the star formation efficiency in the available
gas. After the specific inflow rate has sufficiently decreased --
typically at $z<3$ -- galaxies settle in a self-regulated regime with
efficient star formation. An analytic model shows that this
interaction between infalling gas and young galaxies can significantly
delay star formation and maintain high gas fractions ($>$40\%) down to
$z \approx 2$, in contrast to other galaxy formation models. Idealized
hydrodynamic simulations of infalling gas streams onto primordial
galaxies confirm the efficient energetic coupling at $z>2$, and
suggest that this effect is largely under-resolved in existing
cosmological simulations.
\end{abstract}


\section{Introduction}
\label{sec.intro}
Star forming galaxies at redshift $z \approx 1-2$ have high gas
fractions $\gtrsim$50\%, probed by both their molecular gas and dust
properties (\citealt{daddi10_gasfrac, magdis12, tacconi13}, but note
criticism from \citealt{narayanan12}).
{ Many galaxy evolution models under-predict gas fractions at high
  redshifts} \citep{dutton10,dave12}, and cosmological simulations
find gas fractions as low as 10\% at $z \approx 1-2$ for stellar
masses in the $10^{10}-10^{11}$\,M$_\odot$ range
(e.g. \citealt{ceverino10, keres12}, but higher in some cases,
e.g. \citealt{genel12}).  This likely results from a too-rapid
consumption of gas reservoirs at $z > 2$, where observations suggest
that the efficiency of star formation is more moderate \citep[in
  proportions that remain contentious;][]{daddi07, gonzalez10,
  elbaz11, bouwens12, reddy12, stark13}. The need to delay star
formation in the $z>$2 progenitors of star-forming galaxies,
summarized by \citet{bouche10} and \citet{weinmann12}, may
\citep{henriques13} or may not be solved by stellar feedback --
recently simulated galaxies with stronger feedback still become
star-dominated within their effective radii by $z\approx 3$, with
excessive specific star formation rates at earlier epochs
\citep{ceverino13}. Feedback from active galactic nuclei tends to
exacerbate the problem by lowering gas fractions \citep{dubois12_dual}
if it has a substantial effect at all
\citep[cf.][]{gabor13_bh_growth}. Low metallicity could play a role by
hindering cooling \citep{krumholz12_metal}.  Here we examine another
possibility to maintain high gas fractions: infalling gas streams
could themselves delay star-formation.

Cosmological simulations indicate that galaxies are predominantly fed
by relatively diffuse gas flowing along dark matter filaments, rather
than mergers \citep[e.g.][]{keres05, ocvirk08, brooks09}.  If the
kinetic energy from the infalling gas couples with the galactic disk,
this may increase the turbulent velocity dispersion at early times
\citep[][]{elmegreen10, genel12_analytic}, until grown-up galaxies at
lower redshift manage to self-regulate their gas turbulence.  We
propose that this boosted velocity dispersion will tend to increase
the physical size of the galaxy gas distribution, lowering the gas
density and star formation efficiency.  As we will show, the overall
SF effiency can be reduced by factors of about three at $z \geq 3 $,
making it possible to maintain high gas fractions of 40-50\% down to
$z \approx 2-3$.  We study this process with an analytic model
(\S\ref{sec.model}) that combines the ``bath tub'' approach for galaxy
evolution \citep{bouche10, dave12, lilly13} with analytic descriptions
of star formation in turbulent gas \citep{elmegreen02, krumholz07} and
accounts for the external energy supply.  We also simulate the
hydrodynamic interaction of galaxies with infalling gas
streams (\S\ref{sec.sims}) and confirm that rapid gas infall feeds
only reduced and delayed star formation in $z \approx 3-6$ conditions.

\section{Analytic model}
\label{sec.model}
We develop an analytic model for galaxy evolution that combines three
main physical elements: an estimate of self-regulated
vs. infall-driven gas turbulence, a star-formation rate linked to the
gas density and velocity dispersion, and a standard ``bath-tub''-like
mass budget.

(i) {\bf \em Self-regulated vs. infall-driven gas turbulence: } We use
an approach similar to \citet{elmegreen10} to determine the gas
velocity dispersion.  Internal processes in galaxies that generate
turbulence (e.g. gravitational instabilities, radial flows, star
formation feedback) all saturate about the limit for gravitational
instability, which corresponds to a Toomre parameter $Q=0.7$ for thick
disks \citep{dekel09}. These processes, regardless of their respective
individual contribution, stir the gas disk at a minimal level
$\sigma_{\rm min}=Q \pi G \Sigma_{\rm gas} / \Omega \sqrt{2}$ below
which the gas velocity dispersion cannot drop (in a steady
state). $\Sigma_{\rm gas}$ is the gas surface density, and $\Omega =
v_{\rm rot}/r$ is the orbital frequency of the disk.

The next question is whether external gas infall {\em alone} can stir
the disk above this internally-determined level. Because internal
processes {\em saturate} at $\sigma_{\rm min}$, a contribution from
external sources lower than $\sigma_{\rm min}$ will not add turbulence
-- it will only decrease the relative contribution of internal sources
in the $\sigma_{\rm min}$. External infall at a mass rate
$\dot{M}_{\rm inflow}$ injects kinetic energy at a rate $\dot{E}_{\rm
  turb, infall} = A_{\rm infall} 0.5 \dot{M}_{\rm inflow} v_{\rm
  infall}^2$, where $v_{\rm infall} = \sqrt{2} v_{\rm halo} = \sqrt{2}
(10 G M_{\rm h} H(z))^{(1/3)}$, $M_{\rm h}$ is the halo mass, and
$H(z)$ is the redshift-dependent Hubble parameter
\citep{croton06}. $A_{\rm infall}$ specifies the coupling efficiency
of infall energy to the turbulent energy of the disk.  For simplicity
we assume that $A_{\rm infall}=$ (disk surface area)$/$(stream
cross-sectional area), with a maximum value of 1.  We address the disk
radius and thus the disk area below.  We assume the typical cosmic
stream radius to be $f_{\rm stream}=0.1$ times the virial radius of
the characteristic Press-Schecter halo \citep[cf.][]{dekel06}.  This
typically gives $A_{\rm infall}=1$ at $z\gtrsim2$ for the galaxies we
study here, which is supported by our simulations in \S\ref{sec.sims}.
{  The other key behavior of this coupling is that it becomes weak
  around $z=2$, as required to prevent low-$z$ disks from being too
  thick \citep{genel12_analytic}.  We discuss the coupling further at
  the end of \S\ref{sec.results}.}

While inflows add turbulent energy, dissipation of that energy occurs
over $f_{\rm diss} \approx 3$ disk dynamical times
\citep[cf.][]{maclow98, gammie01}, so that $\dot{E}_{\rm turb, diss} =
E_{\rm turb} / \tau_{\rm diss} = E_{\rm turb} v_{\rm rot}/ (f_{\rm
  diss} R_{\rm gal})$, where $v_{\rm rot}$ is the galaxy rotation
velocity, and $R_{\rm gal}$ is the radius. With the equations we
track the turbulent
energy in the gas, $E_{\rm turb}$, and the velocity dispersion
achievable from infall $\sigma_{infall} = \sqrt 2/3 E_{\rm turb} /
M_{\rm gas}$. Because of the previously mentioned saturation effects,
we assume that the gas settles at a velocity dispersion equal to $\max
(\sigma_{\rm min}, \sigma_{\rm infall})$.

\begin{figure}
\includegraphics[width=84mm]{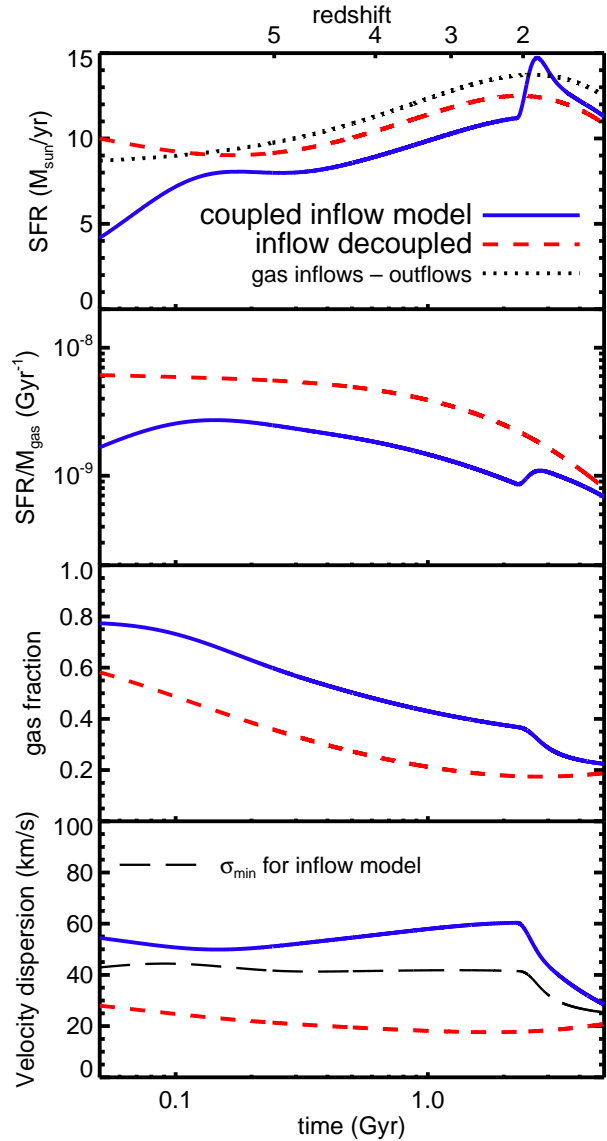}
\caption{In our simple cosmological inflow model (solid blue lines),
  accreting gas suppresses star-formation, allowing a substantial gas
  reservoir to remain in place to $z<2$.  We compare this fiducial
  model with an otherwise identical model where inflowing cold streams
  have no effect on the galaxy (other than to provide fuel for SF;
  dashed red lines).  {\bf Top:} The SFR is suppressed by $\sim$25\%
  due to infalling gas at high redshifts.  We also show net gas
  inflows to the galaxy for comparison (dotted line).  {\bf Second
    panel:} Global galaxy SF efficiency, SFR$/M_{\rm gas}$, is reduced
  by a factor of $\sim 3$ due to inflows.  {\bf Third panel:} Gas
  fraction.  By suppressing the SFR, infalling cold streams allow the
  gas fraction to remain high.  {\bf Bottom:} Velocity dispersion.
  For the coupled-inflows model, we also show $\sigma_{\rm min}$
  (long-dashed line). Infalling cold streams increase the velocity
  dispersion by up to $\sim$50\% compared to $\sigma_{\rm min}$, and a
  factor of 3 above the control simulation (which has a lower gas
  fraction).}
\label{fig.time_series}
\end{figure}

(ii) {\bf \em The star formation rate} is determined based not just on
  the gas density, but also on its velocity dispersion. We assume the
  gas takes on a log-normal density probability distribution function
  (PDF) typical of supersonic turbulence \citep[e.g.][]{padoan97},
  independent of the main source of turbulence. The peak of the
  log-normal is determined from the average gas density of the galaxy (see below)
and the width of the PDF is $(\ln(1+0.75
  (\sigma/c_s)^2))^{(1/2)},$ where the sound speed $c_s$ is assumed to
  be $\approx 10$\kms for the star-forming gas \citep{padoan97} 
We assume that
  gas at number densities above 100 cm$^{-3}$ forms stars at a
  volumetric rate of $\dot{\rho}_* = \epsilon_* \rho / t_{\rm ff} =
  \epsilon_{*} (32 G \rho^3 / (3 \pi))^{(1/2)}$, where $t_{\rm ff}$ is
  the free-fall time and $\epsilon_*=0.01$ is the local star-formation
  efficiency per free-fall time \citep[cf.][]{krumholz12_sflaw}.
Numerical integration over the density PDF and galaxy volume gives the
total SFR of the galaxy.

Determining the average gas density requires an estimate of the galaxy
geometry.  {  The disk radius depends on the galaxy's angular
  momentum, which is highly uncertain for high-$z$ galaxies.  For
  simplicity and consistency with previous work,} we follow
\citet{krumholz12_metal} to obtain the standard, undisturbed disk
scale radius of $R_{\rm scale} = 0.5 \lambda R_v,$ where $\lambda$ is
the halo spin parameter and $R_v$ is the halo virial radius.  We use
$\lambda=0.07$ as a typical value \citep{dutton11}.  This radius is
increased when infalling gas increases the velocity dispersion and
scatters gas to larger radii. We can estimate this effect by taking
$v_{\rm rotation}^2 \approx v_{\rm circ}^2 - 3\sigma^2$ for the
rotational specific energy, then assuming angular momentum is
approximately conserved with $R_{\rm scale} v_{\rm circ} \approx
R_{\rm disturbed} v_{\rm rotation}$. Here $v_{\rm circ}$ is the
circular velocity and $v_{\rm rotation}$ is the actual rotational
velocity.  The disk scale radius becomes $R_{\rm disturbed} \approx
R_{\rm scale} (1.0 - 3\sigma^2 / v_{\rm circ}^2)^{(-1/2)}$.  For
simplicity, we multiply this scale radius by 1.7 to obtain the total
disk radius, as appropriate for exponential disks.  The scale height
of the gas disk is given by $H = \sigma^2 / (\pi G \Sigma_{\rm s+g})$,
where $\Sigma_{\rm s+g}$ is the total mass surface density of stars
and gas in the disk.  Thus, when inflowing gas streams increase
$\sigma$, both the radius and scale height increase.  The average gas
density decreases because the rapidly infalling gas does not
immediately settle into a thin, self-gravitating disk.

(iii) {\bf \em ``Bath tub''-like mass budget: } Cosmological inflow
acts as a gas source, while star formation and galactic winds act as
sinks.  We take the cosmological mass inflow rate of gas into a dark
matter halo of mass $M_{\rm h}$ at redshift $z$ to be $\dot{M}_{\rm
  inflow} = 6.6 f_b (M_{\rm h}/10^{12})^{1.15} (1+z)^{2.25}$
\citep{dekel09}.  We assume that the cosmological fraction of baryons
$f_b=16.5$ per cent accompanies the dark matter into the halo, and for
simplicity we assume the baryons all enter as gas via cold streams
(rather than mergers).  {  Although recent simulations raise doubt that
cold streams can penetrate the halo \citep{nelson13}, they are likely
the dominant mode of galaxy fueling in low mass halos at high-$z$.}
Star-formation (as described above) removes gas from the galaxy's
reservoir, and we assume that young stars drive galactic winds that
eject gas from the galaxy at a rate equal to the SFR.

Using these equations to track the turbulent energy, star-formation,
and gas reservoir, we evolve a model galaxy over cosmic time with a
simple numerical integration scheme.  To make the discussion concrete,
we focus on a Milky Way progenitor galaxy with an initial halo mass of
$5\times10^{10} M_{\sun}$ at $z=6$ and a final ($z=0$) halo mass of
$1.7\times10^{12} M_{\sun}$.  The galaxy has an initial baryonic mass of
$2.5\times10^{9} M_{\sun}$, an initial gas fraction of 75 per cent, and an
initial velocity dispersion of 50~\kms, but the galaxy rapidly evolves
towards an equilibrium independent of these choices.

\begin{figure*}
\includegraphics[width=168mm]{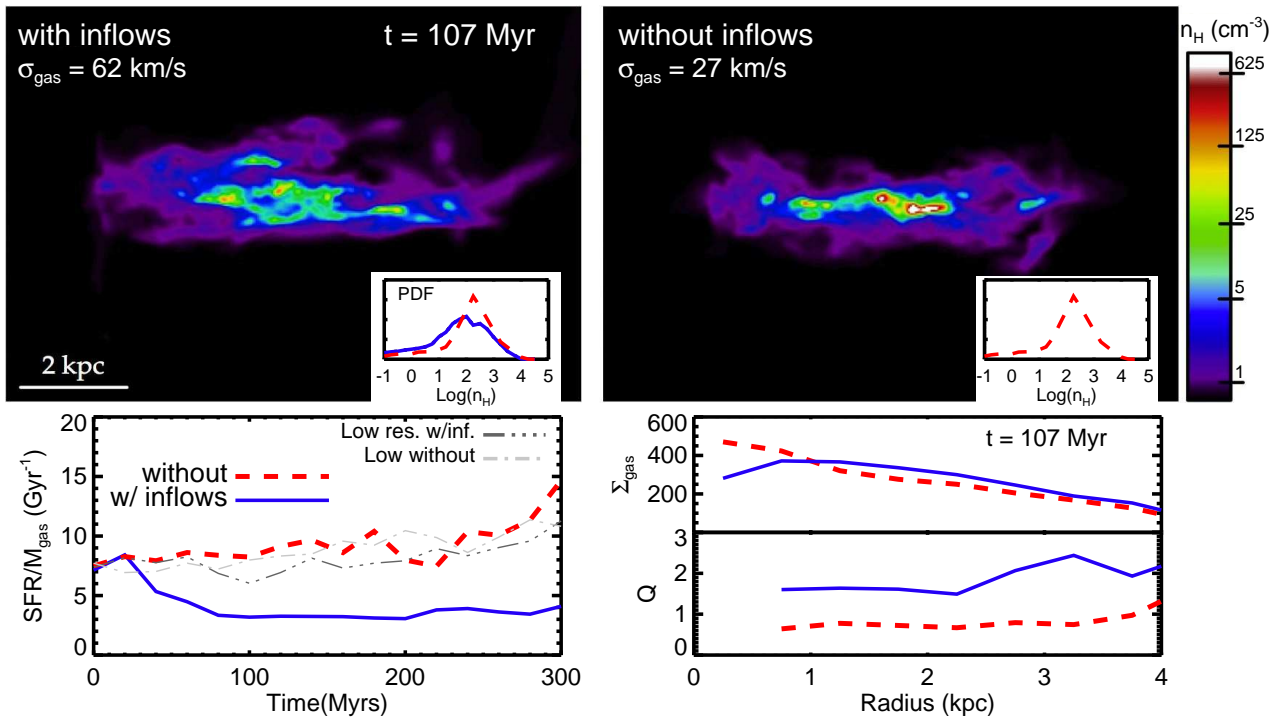}
\caption{{\bf Top:} Gas density images of simulations of a $z\sim5$
  isolated galaxy with (left) and without (right) cold inflows, at a
  time $t=107$~Myr after the cold streams meet the disk.  Cold inflows
  increase the gas velocity dispersion and scale height of the galaxy.
  Inflows also lead to a (normalised) density PDF (inset panels) with
  less gas at high, star-forming densities (i.e. $>10^2$~cm$^{-3}$).
  {\bf Bottom left:} The SF efficiency (SFR$/M_{\rm gas}$) is $2-3$
  times lower with inflows (blue solid lines) than without (red
  dashed).  We also show results for low-resolution simulations with
  and without inflows (gray dot-dashed lines) -- without sufficient
  resolution, the SFE is unaffected by inflows.  {\bf Bottom right:}
  Radial profiles of gas surface density $\Sigma_{\rm gas}$ (in
  \Msun~pc$^{-2}$) and Toomre $Q$ parameter.  Inflows lower the gas
  surface density within the central regions, and increase the
  stability to $Q>1$ across the disk.}
\label{fig.disk_image}
\end{figure*}

\subsection{Results of the inflow model}
Figure \ref{fig.time_series} compares the results of our fiducial
coupled-inflow model with a control model, in which inflow is
decoupled from the disk.  In the decoupled model, the galaxy has the
same mass inflow rate, but we neglect any energy coupling between inflows
and the disk.

At $z>2$ in our infall model (solid blue lines), inflows are strongly
coupled to the galactic disk.  They deposit turbulent energy into the
disk, causing a boost in the velocity dispersion (bottom panel).  The
boosted velocity dispersion slightly lowers the average gas density of
the galaxy, thus lowering the global SF efficiency and SFR.  Because
the SF efficiency is suppressed, gas is allowed to accumulate in the
galaxy, leading to a sustained high gas fraction (higher by a factor
of $\sim2$).  The specific SFR remains almost unchanged compared to the
control model because both the SFR and stellar mass are suppressed by infall.

At $z=2$ the coupling between the infalling cold streams and galaxy
disk becomes less efficient {  (with an abruptness owing to our simple
assumptions about $A_{\rm infall}$)}.  This occurs when the typical
filament size equals the size of the galaxy disk.  When the coupling
weakens, the turbulent energy in the disk decays, causing the velocity
dispersion to decline (bottom panel near $z=2$).  This leads to an
increase in gas density that causes a brief spike in the SFR (top
panel).  The gas fraction begins to decrease and eventually the SFR
does as well, and all quantities converge toward the ``decoupled
inflow'' model at low redshifts.  The galaxy converges towards its
normal state as a self-regulated, star-forming disk.


In summary, the main effect of infalling gas is to increase the
velocity dispersion, ``puff up'' the galaxy disk, lower the
SFR and SF efficiency, and increase the gas fraction.  These effects
persist as long as the coupling between infalling gas and the galaxy
disk is strong (as we have assumed).  When the coupling becomes weak
(at $z\approx2$ in our model), the galaxy converges towards the
solution where infalling gas is decoupled from the disk.


\section{Simulations}
\label{sec.sims}
To test our model, we simulate the accretion of cold gas flows by
young galaxies with the Adaptive Mesh Refinement (AMR) code
\textsc{ramses} \citep{teyssier02}. A key difference with existing
idealized and cosmological simulations is that we ensure relatively
high resolution in low-density gas, at the expense of the maximal
resolution in the densest regions. We start with a coarse grid of
resolution 1.0\,kpc, and refine each cell into eight smaller cells
when its gas mass exceeds 200\,$M_{\odot}$ and/or it contains more
than 240 particles (of stars and/or dark matter). The maximal
resolution of 32\,pc is ensured as soon as the gas number density
exceeds 0.2\,cm$^{-3}$ (with 64\,pc resolution at
0.025\,cm$^{-3}$). We allow cooling down to $10^4$~K and enforce a
pressure floor to keep the Jeans length resolved by at least four
cells.

{  Old stars, new stars and dark matter are modeled with particles of
$2\times 10^3$, 800, and $8\times 10^3$\,M$_{\odot}$,
respectively.} Star-formation occurs in gas denser than 1\,cm$^{-3}$,
with an efficiency of 0.01 per local gravitational free-fall time
\citep[cf.][]{zuckerman74}. The low density threshold
ensures that variations in the SFR are not caused by
artefacts near the highest resolvable densities, although in practice
most of the SFR is obtained from gas at and above 100\,cm$^{-3}$. We
include stellar feedback -- photoionization and radiation pressure,
using the \citet{renaud13} model with a trapping parameter $\kappa =
5$, and supernovae feedback: ten Myrs after its birth, a star particle
dumps thermal energy in the nearest gas cell, at a rate of
$10^{51}$~ergs for each $10M_{\sun}$ of progenitor mass. We account
for non-thermal processes in supernova remnants by decreasing the gas
cooling rate in the affected cells, as proposed by
\citet{teyssier13}. This can also be seen as a way to prevent
over-cooling of the thermal phases in supernovae remnants
\citep[cf.][]{stinson06}.

We use two galaxy models, roughly representative of progenitors of
Milky Way-mass galaxies at redshifts $z \approx 5$ and $z \approx
1-2$. They start with a stellar mass of $1.3 \times 10^9$\,M$_{\odot}$
for $z \approx 5$ (respectively $8.0 \times 10^{10}$\,M$_{\odot}$ for
$z \approx 1-2$) and a gas fraction of 75 per cent (resp. 50 per
cent). The disk scale-length is 1.8\,kpc (resp. 4\,kpc) and a bulge of
radius 240\,pc (resp. 500\,pc) contains 15 per cent of the stellar
mass. Dark matter halos with NFW profiles are used with concentration
parameters set to 7.0. {  A hot ($>$10$^6$K) atmosphere is included
  with an homogeneous density (a few $\times 10^{-5}$~cm$^{-3}$) 
such
  that the mass comprised in the dark halo virial radius is 37\% of
  the disk mass.}

Each galaxy is modeled with and without accreting cosmological
streams. The inflow rate is 15\,M$_{\odot}$~yr$^{-1}$ for the
$z\approx 5$ case and 80\,M$_{\odot}$~yr$^{-1}$ for the $z \approx
1-2$ one -- for the lower-mass, higher-redshift progenitor, this
corresponds to a twelve times higher specific inflow rate at $z
\approx 5$, as expected from theory (cf. \S\ref{sec.model}). The inflowing gas
is introduced in three streams. {  The streams originate at a distance
from the galaxy center equal to 1.2 times the virial radius of the
dark matter halo. They have a circular cross-section with an azimuthal
profile as follows: a central density set by the required inflow rate,
an exponential decay with a scale-length equal to the galaxy disk
radius, and a truncation at 2.5 times this scale length.} In the disk
frame, the streams are set to cross the virial radius at latitudes of
(+35, -45 and +65) degrees, respectively, and longitudes of (-90,+90
and +180) degrees, respectively. They contain respectively 45, 35 and
20 percent of the mass inflow rate, have impact parameters of 1.5, 2.0
and 1.5 disk scale lengths with prograde orientations except for the
third and weakest one. This configuration, while arbitrary, is
consistent with cosmological simulations \citep[cf.][]{danovich12},
and so is the gas density along the flow axis ($0.01-0.05$~cm$^{-3}$,
$\sim10^3$ times lower than the average initial density in the galaxy).

\subsection{Simulation results}
\label{sec.results}
\begin{figure}
\includegraphics[width=84mm]{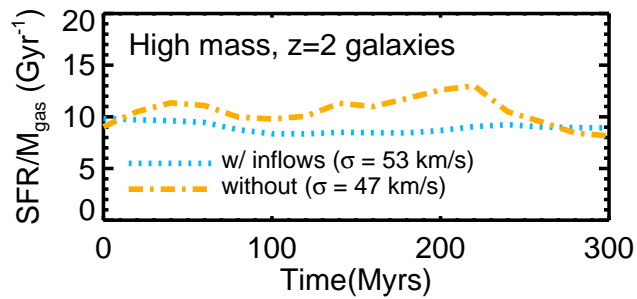}
\caption{SF efficiencies vs. time for our two massive $z=2$
  galaxy simulations.  In this case, the effect of cold inflows is weak,
  typically changing the SF efficiency by $\sim20$ per cent or less.  }
\label{fig.sfe_highmass}
\end{figure}
We simulate each galaxy for $\approx2$ rotation periods ($300+$~Myr).
Figure \ref{fig.disk_image} shows edge-on gas maps of the two low-mass
$z\sim5$ galaxies at $t=107$ Myrs.  The cold inflows puff up the disk,
leaving the galaxy with fewer regions at the highest densities.  { 
  Inflows increase the effective scale height from 325~pc to 612~pc,
  and the half-mass radius from 2.12~kpc to 2.86~kpc.}  We measure
mass-weighted gas velocity dispersions and star-formation rates within
5 kpc of the galaxy center, {  along with radial profiles of gas
  surface density and an estimate of the Toomre Q parameter} (assuming
the gas and stellar velocity dispersions are equal).  The velocity
dispersion in the inflows simulation is higher by a factor of $\sim2$,
leading to a thicker disk and a SF efficiency lower by a factor of
$\sim2-3$.  The figure also shows the (volume) density PDFs of the two
simulations, illustrating that cold inflows can reduce the amount of
gas at high, star-forming densities above 100~cm$^{-3}$.  {  The
  lower right panels show that the central gas surface density,
  $\Sigma_{\rm gas}$, is actually decreased due to inflows.  Rather
  than merely adding gas to the galaxy, the inflows contribute energy
  that stabilises the disk at all radii, as indicated by the boosted
  Toomre Q parameter.}

Thus abundant, filamentary cold gas inflows can disturb small galaxies
at $z\sim5$, boosting the velocity dispersion and effectively lowering
the SFR.  For more massive galaxies at lower redshifts, the cold
streams do not affect the gas disk as strongly, as demonstrated by our
$z\sim2$ simulations.  In these simulations the velocity dispersion is
barely changed (53~\kms with inflows versus 47~\kms without), and the
SF efficiency remains within $\sim20$ per cent, as shown in Figure
\ref{fig.sfe_highmass}.  This agrees qualitatively with the
`high-redshift' simulations of \citet{hopkins13_turb}, which are
broadly representative of massive $z\sim2$ galaxies.  We speculate
that the $\sim 10\times$ higher specific inflow rate ($\dot{M}_{\rm
  inflow} / M_{\rm gas}$) in our low-mass $z\sim5$ galaxies leads to a
stronger coupling between infalling gas and the disk.  Cosmological
simulations -- even zoom-ins -- generally lack the resolution in the
low-density circum-galactic gas to model this effect accurately
\citep[although they do show a messy, rapidly varying connection
  region between filaments and the galaxy, e.g.][]{danovich12,
  dubois12_feeding}.

{  These simulations imply that energy from infalling streams can,
  at least under some circumstances, couple strongly to the galactic
  disk.  In our analytic model (\S\ref{sec.model}) we posit that the
  coupling ($A_{\rm infall}$) is strong at $z>2$ and then weakens as
  the streams become larger than the galaxy.  Our simulations, though
  limited in the scope of parameter space, suggest instead a
  dependence of the coupling on halo mass or specific accretion rate.
  This dependence also leads to reduced SF efficiency at high-$z$
  followed by a transition to normal efficiency at lower $z$.  Another
  possibility is that cold streams are disrupted before they fully
  penetrate the halo \citep{nelson13} -- while such disruption
  precludes interaction with the disk, it could effectively delay
  star-formation by prolonging the time spent in the circum-galactic
  reservoir.  In our simulations the streams penetrate an idealised
  hot atmosphere without disruption, but cosmological simulations with
  high resolution in a more realistic circum-galactic gas distribution
  may be needed to confirm this result.  Although some physical
  details remain in question, our model and simulations support a
  scenario where infalling cold streams delay star formation in
  low-mass galaxies at high-$z$, but have little effect on their more
  massive descendents at $z\lesssim2$.  }

\section{Conclusion} 
\label{sec.summary} 
Our simple model of cold gas inflows into high-redshift galaxies shows
that if the coupling between inflows and the gas disk is strong, then
inflows can lower the star-formation efficiencies by factors of 3 and
keep gas fractions above 40 per cent until $z<2$.  While previous
models suggest the coupling between inflows and the disk is fairly
weak \citep[e.g.][]{klessen10, hopkins13_turb}, our simulations with
inflows show that this coupling can be strong if enough gas is
inflowing.  When the coupling is strong, cold flows don't settle into
a self-regulated disk, but rather stir up the disk and suppress
star-formation.  Our work implies that in the high-$z$ Universe where
galaxies are small and inflow rates are large, the energetic injection
of cold inflows has an important impact on galaxy evolution.


 
 

\section*{Acknowledgements}
We thank Romain Teyssier for making \textsc{ramses} publicly
available; the referee for prompt and thoughtful comments; and Daniel
Ceverino, Avishai Dekel, and Florent Renaud for useful discussions.
We acknowledge support from the EC through grants ERC-StG-257720 and
the CosmoComp ITN.  Simulations were performed at TGCC and as part of
a GENCI project (grants 2011-042192 and 2012-042192).

\bibliographystyle{mn2e} 

\bibliography{paper}


\label{lastpage}

\end{document}